\magnification=1200
\def\ssp{\baselineskip=12pt plus 1 pt minus 1 pt}

\ssp
\def\title#1{\centerline{\bf #1}}
\def\author#1{\bigskip\centerline{#1}}
\def\sec#1{\bigskip\centerline{{\bf #1}}\bigskip}
\def\subsec#1{\medskip\centerline{{\it #1}}\medskip}

\def\ni{\noindent}

\def\cm{{\rm cm}}

\def\erg{{\rm erg}}
\def\K{{\rm K}}

\def\Jy{{\rm Jy}}

\def\kpc{{\rm kpc}}
\def\Mpc{{\rm Mpc}}
\def\km{\rm km }

\def\s{{\rm s} }
\def\D{\Delta}

\def\v#1{{\bf#1}}

\def\eg{{\it e.g.~}}
\def\ie{{\it i.e.~}}
\def\etal{{\it et al.~}}
\def\etc{{\it etc.~}}
\def\ni{\noindent}

\def\msun{M_\odot}
\def\Lya{Ly$\alpha$~}
\def\gapprox{\lower.4ex\hbox{$\;\buildrel >\over{\scriptstyle\sim}\;$}}
\def\lapprox{\lower.4ex\hbox{$\;\buildrel <\over{\scriptstyle\sim}\;$}}
%%%
\title {\Lya Line Formation in an Expanding H~I Superbubble}
\title {in the Primeval Galaxy DLA~2233+131} 
\author { Sang-Hyeon Ahn$^1$ and Hee-Won Lee$^{1,2}$}
\author{1. Dept. of Astronomy, Seoul National University}
\author{2. Research Institute for Basic Sciences, Seoul National University}
\author{Seoul, Korea}
\author{e-mail : sha@astro.snu.ac.kr, hwlee@astro.snu.ac.kr}

\sec{Abstract}

%%Intro and Method
Various types of galaxies observed in the cosmological scales show
P-Cygni type profiles in the \Lya emission lines. 
The main underlying mechanism for the profile
formation in these systems is thought to be the frequency redistribution
of the line photons in an expanding H~I component surrounding
the emission source.

A Monte Carlo code is developed to investigate the \Lya line transfer
in an optically thick and moving medium with a careful consideration
of the scattering in the damping wing. Typical column densities and
the expansion velocities investigated in this study are 
$N_{HI}\sim 10^{17-20}~\cm^{-2}$ and $\Delta V\sim 100~\km~\s^{-1}$.

The main features in emergent line profiles include
a primary emission peak in the red part and a much weaker
secondary peak in the blue part. We investigate the dependence of 
the profile on the kinematics and on the column density.
The primary peak recedes to the red and the width of the feature increases 
as $N_{HI}$ increases. 
It is also found that symmetric double peak profiles
are obtained in the static limit, where bulk motion is negligible.

The P-Cygni type profile in the \Lya emission line of DLA~2233+131
($z=3.15$) is noted to be similar to those found in the nearby
galaxies and distant ($3<z<3.5$) H~II galaxies.
Our numerical results are applied to show that
the DLA system may possess an expanding H~I supershell
with bulk flow of $\sim 200~\km~\rm s^{-1}$ and that the H~I column density
$N_{HI}$ is approximately $10^{20}\cm^{-2}$. 

%%%%SHA%%
From the observed \Lya flux and
adopting a typical size of the emission region $\sim 1~\kpc$,
we estimate the electron density of the H~II region
to be $\sim 1~\cm^{-3}$ and 
the mass of H~II region $\sim 10^8~\msun$. We also conclude that
it requires $\sim 10^3$ O5 stars for photoionization, which is
comparable to first-ranked H~II regions found in nearby spiral 
and irregular galaxies.

We briefly review the physical quantities of several
astronomical objects which may possess similar kinematic
properties to those of DLA~2233+131. We point out 
the fact that this kind of outflowing media are often
accompanied by the sites associated with active star formation.


\vfill \eject
\sec{1. Introduction}

A typical high-redshift quasar spectrum shows a variety of absorption 
features blueward of the broad \Lya emission line, most
of which are not associated with the quasar itself. 
These absorption systems are thought to be related with the neutral
hydrogen and metal elements in the intervening medium along the line
of sight to the quasar, and therefore constitute an ideal subject for
studying the cosmological structure formation processes.
Depending on the neutral hydrogen column density $N_{HI}$,
the absorption systems are usually classified into three classes,
$\ie$ damped \Lya (hereafter DLA) systems, 
Lyman limit systems, and \Lya forest systems.

With the advent of the Hubble Space Telescope and 10~m class telescopes
such as Keck I, II, detailed analyses of absorption line profiles
have been performed, and provided very important kinematic 
information about the absorption systems.

Recently Prochaska \& Wolfe (1997) analyzed the absorption line profiles of
the 17 DLAs obtained with the HIRES eschelle spectrograph on the Keck 
telescope and concluded that the kinematics of the DLAs
supports the hypothesis that they are rapidly rotating ``cold'' 
disks (see also Haehnelt \etal 1997). 

There have been a number of DLA galaxy candidates possibly exhibiting 
\Lya emission features in their spectra (\eg Hunstead \etal 1990, 
M\"oller \& Warren 1993, Pettini \etal 1995 \etc).
However, the first clear \Lya emission feature was detected 
in the spectra of the DLA~2233+131 (Djorgovski \etal 1996).

Djorgovski \etal (1996) reported an isolated galaxy 
(or a protogalaxy) 
$2.3"$ (or 17 kpc) away from the line of sight to the quasar, QSO 2233+131 
by detecting the \Lya emission having the same redshift as DLA 2233+131. 
They also noted that the emission profile is asymmetric to the red and
that there is a possible absorption feature to the blue. The width of the
absorption feature is measured to be $\sim 200~\km~\s^{-1}$.
It is not certain that the absorption is related to a very near \Lya forest,
as was pointed out by several researchers (Djorgovski \etal 1996,
Lu \etal 1997). 
However, the similarity of the profile to the P-Cygni profile leads to 
an interesting possibility that an expanding medium envelops 
the emission source, which can be thought to be a supershell
seen in nearby galaxies (see section 4. of Irwin (1995) for a review 
of outflowing structures).
Since the first firm identification of DLA candidate DLA~2233+131, 
Djorgovski \etal (1997) found several new candidates.
%%%%%%%%%%%%%%%%%%

In addition to this direction, there are at least two other groups
adopting efficient programs to find out primeval galaxies.

Firstly, Giavalisco \etal (1996) found that 
there are a number of star-forming galaxies at $3.0<z<3.5$
possessing compact spheroidal cores often surrounded by low surface
brightness nebulosities. Morphologically similar objects are also 
found in the Hubble Deep Field (Steidel \etal 1996a).
Moreover, Steidel \etal (1996b) argue that
the spectra of galaxies with $z>3$ are remarkably similar to those of
nearby star-forming galaxies and often show \Lya lines with P-Cygni profiles
characteristic of expanding H~I media.
The H~I column density is inferred to be $10^{20}~\cm^{-2}$ in these
systems.

One excellent example of nearby star-forming galaxies 
is provided by Legrand \etal (1997) and
Lequeux (1995) who performed a detailed analysis of the emission profiles 
of the star-forming galaxy Haro 2.
They suggest that these P-Cygni type profiles are formed 
by the expanding H~I shell surrounding an H~II region.
At least three more galaxies showing such profiles are found 
in the literature (Kunth \etal 1996).

%%%%%%%%%%%%
Recently Steidel \etal (1997) used the Lyman break criterion 
to obtain $\sim 80$ 
new candidates that are similar to those found previously.
What is interesting is that $\sim 75\%$ of them exhibit P-Cygni
type \Lya profiles (Pettini \etal 1997).
Lowenthal \etal (1997) also observed the spectra of 24 remote 
galaxies selected from the Hubble Deep Field.

%%%%%%%%%%%
Secondly, using the flux magnification via
gravitational lensing by foreground clusters of galaxies,
very faint and remote galaxies can be found.  
Trager \etal (1997) found a multiply-imaged 
galaxy at $z\sim4$ gravitationally lensed by the rich cluster CL0939+4713,
which are the same kind of those discovered by 
Lowenthal \etal (1997). Using the same tactics, Franx \etal (1997)
discovered the currently remotest galaxy at $z=4.92$. Recently Frye 
\etal (1997) found the remote galaxy at $z=4.04$ which is lensed by the
cluster of galaxies, A2390 ($z=0.23$).
Surprisingly, $\sim 75\%$ of the primeval galaxies described above
exhibit P-Cygni type \Lya emission lines, which implies the ubiquity
of an expanding HI envelope around HII region.
%%%%%%%%%%%

So far, many investigations about the P-Cygni profiles have been concentrated
on the outflowing systems possessing rather moderate optical depth.
The Sobolev approximation is regarded as a powerful method to understand
the behavior of resonantly scattered photons in an expanding medium
with a bulk flow much larger than the thermal speed 
(Sobolev 1960, Mihalas 1978). 
However, when the medium has high optical thickness,
the scattering in the damping wing is not negligible any more
and the validity of the Sobolev approximation
becomes questionable. 

In this paper, we develop a Monte Carlo code for  
the radiative transfer of \Lya photons 
in an optically thick and expanding medium 
that is expected in the H~II superbubbles in star-forming galaxies.
Extrapolating the kinematical information from our line profile study,
we try to find a possible relation of 
the P-Cygni type absorption profile to the existence of 
superbubble in the primeval starburst galaxies.

The paper is composed as follows.
In section 2 we describe the Monte Carlo method 
to deal with the \Lya transfer in a thick moving medium.
In section 3 we present the main results.
We apply our results to the absorption features 
in the \Lya emission of DLA 2233+131 in the following section.  
Discussions and the implications for the identity of 
DLA~2233+131 are given in the final section.

\sec{ 2. Monte Carlo Simulations }

\sec{2.1 Model Description}

In this subsections we give a brief model description 
and a basic atomic physics for the Monte Carlo 
code, which computes the profile of the emergent \Lya photons scattered 
inside an optically thick and expanding medium.

The photon source is assumed to be located at the center of the 
spherical scattering medium which is truncated at $r=r_{max}$. 
We assume that the constant density distribution $n(r) = n_0$
throughout the scattering region. 
We only consider an isotropic and point-like
line emission source in this work, which is not a bad approximation for
emission sources such as H~II regions surrounded by H~I regions.

The velocity field of the scattering medium is assumed to be given
by a Hubble-type flow, \ie, 
$$ 
\v v = H_v \v r, 
\eqno(1)
$$
where $H_v$ is the bulk flow velocity gradient,
$\v v$ is the bulk velocity, and $\v r$ is the distance from the center
of the velocity field. 
In this type of velocity field, the distance is more
conveniently measured by a parameter $s$ defined by
$$
s \equiv {H_v r / v_{th}},
\eqno(2)
$$
where $v_{th}$ is the thermal velocity.

If we introduce the Doppler width $\D \nu_D$ defined by
$$
\D \nu_D \equiv \nu_0 {v_{th} \over c} ,
\eqno(3)
$$
then the frequency of a line photon is also conveniently described by the
normalized frequency shift $x$ defined by
$$x\equiv (\nu-\nu_0)/\D\nu_D. \eqno(4)$$

Because the scattering medium is assumed to be optically thick,
scattering in the damping wing should be considered with much caution.
Therefore we introduce the damping coefficient, $a$, 
normalized by $\D \nu_D$,
\ie,
$$
a = \Gamma / 4\pi \D \nu_D ,
\eqno(5)
$$
where $\Gamma$ is the damping constant. The scattering cross section
in the rest frame of the scatterer is then given by
$$\eqalign{
\sigma_{\nu} &=\sigma(x)= {\pi e^2 \over m_e c} f_{osc} 
{\Gamma/4\pi^2 \over (\nu-\nu_0)^2+(\Gamma/4\pi)^2} \cr
 &= {\pi e^2 \over m_e c} f_{osc} {1\over \pi\D\nu_D}{a\over x^2+a^2}, \cr}
  \eqno(6)
$$
where $f_{osc}$ is the oscillator strength, $m_e$ is the electron mass
(Rybicki \& Lightman 1979).

The scattering optical depth $\tau_{12}(x)$ of a line photon with 
frequency shift $x$ and wave vector $\v k_i$ corresponding to the 
distance from the position
$\v r_1$ to $\v r_2$ is computed to determine the subsequent scattering
position in the Monte Carlo calculation. Due to the Hubble-type velocity
field and a constant density field inside the scattering medium,
we get the same Hubble-type flow if we make a translation so that $\v r_1$
coincides with the new origin. 
The line center frequency $\nu'_0(r)$ in the observer's frame differs from 
that in the scatterer's rest frame by the Doppler shift, which is given by
$$\nu'_0(r)=\nu_0(1-H_v r/c+v_{loc}/c), \eqno(7)$$
where $r$ is the distance from the new origin $\v r_1$ and $v_{loc}$ is
the local velocity of the scatterer due to the thermal motion.

The normalized frequency shift $x'(r)$ is now simply given by
$$x'(r) = x-s+u, \eqno(8) $$
where $u\equiv v_{loc}/v_{th}$.
Therefore the scattering cross section $\sigma(x)$ becomes a function
of the position due to the coupling of the line center frequency and
the position determined by the expansion of the medium.

The thermal distribution of the scatterers is assumed to be given by
a Maxwellian distribution, that is,
$$
n=n_0\int dv_{loc}\ f(v_{loc}), 
\eqno(9)$$
where $f(v_{loc})={1\over \sqrt{\pi}v_{th}} e^{-({v_{loc} \over v_{th}})^2}
= {1\over \sqrt{\pi}v_{th}} e^{-u^2}$ is the Maxwellian distribution
function.

Combining Eqs.~1-9, we obtain
$$ \eqalign{
\tau_{12}(s) &=\int_{-\infty}^{\infty}du \int_0^{s}\ ds'\  n_0\ f(u) 
\sigma[x'(s')] \cr
& = \tau_0 \int_{-\infty}^{\infty} du \ e^{-u^2}
\left[ \tan^{-1}\left( {{u+x} \over a} \right) - \tan^{-1}\left(
{u+x-s \over a} \right)\right] .\cr} 
\eqno(10)
$$
Here, $\tau_0$ is the Sobolev-type optical depth defined by
$$ 
\tau_0 = { {\pi e^2} \over {mc}} f_{osc} 
{n_o \over \pi^{3/2}} {\lambda_0 \over H_v}
\eqno(11)
$$
(Sobolev 1960).

The inverse transformation of Eq.~10 is obtained to find 
the normalized path length $s$ corresponding
to a given scattering optical depth $\tau_{12}$.
The escape optical depth $\tau_{esc}$ is also obtained by
setting $s=s_{max}$, where 
$$
s_{max}=H_v r_{max}/v_{th}=\D V_{bulk} /v_{th}.
\eqno(12)
$$
Here, $\D V_{bulk}$ is the difference of the bulk velocities 
at the center and the boundary of the scattering medium.

The conversion of the H~I column density and the Sobolev type
optical depth $\tau_0$ is then given by
$$
\tau_0 =10^4\ 
\left({N_{HI} \over 4.1 \times 10^{18}~\cm^{-2}}\right) 
\left( {H_v \cdot r_{max} \over 100~\km~\s^{-1}} \right)^{-1}
 \left( {f_{osc} \over 0.4162} \right) . \eqno(13)
$$

\sec{2.2 Monte Carlo Approach}

In this subsection, we present a detailed description of 
the Monte Carlo procedure.
There are a few approaches to the resonance
line transfer in an optically thick and static medium (Osterbrock 1962, 
Adams 1972, Harrington 1973, Sengupta 1994, Gould \& Weinberg 1996).
However, there are also studies on the line formation 
in a thick expanding medium (\eg Rybicki \& Hummer 1978).

The Sobolev approximation has been one of the most favored methods
in dealing with the line transfer in an expanding medium.
However, the validity of the Sobolev method is limited to the
cases where the bulk flow is much larger than 
the thermal velocity and the optical depth of the medium
does not greatly exceed unit. This is because the scattering
in the damping wing is no more negligible in an optically
thick medium where the typical line center optical depth 
$\tau_c \gapprox 10^4$.
Therefore we develop a Monte Carlo code
in which the scattering in the damping wing is carefully 
considered in a thick moving medium.

The Monte Carlo code begins with the choice of the frequency 
and the propagation direction $\v k_i$ of the incident photon 
from an assumed \Lya profile. 

Then we determine the next scattering site 
separated from the initial point by the normalized propagation length 
$s$ defined by Eq.~2, which corresponds to the optical depth of
$\tau=-\ln (\it \v R)$, where $\it \v R$ is a uniform random number in the
interval $(0,1)$. Due to the unwieldiness of the inverse relation of
Eq.~10, we tabulate the normalized path length $s(x, \tau)$ in advance
and look up the table to find $s$ by the linear interpolation. For our
numerical computation, the range of $x$ is taken to be
from $-4s_{max}$ to $4s_{max}$ with a step of $\D x =0.5$, and $\tau$ 
runs from $0$ to $4s_{max}$ with a step of $0.002$, where
$s_{max}=10$ is assumed.

In Fig.~1 we present the $s(x,\tau)$ table as a surface plot.
We restrict the propagating length $s$ to be $0<s<2s_{max}$, and
set $s=2s_{max}+1$ if the true value of $s(x,\tau)$ exceeds $2s_{max}$.
A U-shaped distribution of $s$ is seen in the figure, which
is naturally explained by the fact that the scattering medium is
transparent for extreme blue and red photons.

The emitted photon traverses a distance $s$ found by 
the above procedure and is scattered off if $s<2s_{max}$.
In this scattering event 
the frequencies of the absorbed photon and the re-emitted one
in the rest frame of the scatterer should be matched. 
For the case of an optically thin and expanding medium, 
this frequency matching is combined with the Sobolev approximation to yield
the absorption profile in the form of the Dirac delta function.

As mentioned earlier, in contrast with the case of a thin medium, 
for a very thick medium
the scattering in the damping wing is not negligible.
A good care need to be exercised to distinguish the scattering
in the damping wing from the resonance scattering,
because they show quite different behaviors in the properties 
including the scattering phase function and the polarization
(Lee 1997, Lee \& Blandford 1997, Blandford \& Lee 1997).

Because the natural line width is much smaller than the Doppler width,
the local velocity of the scatterers that can resonantly
scatter the incident photon is practically a single value.
However, when the scattering occurs in the damping wing, 
the local velocity of the scatterer may run a rather large range.
Therefore, in order to enhance the efficiency of the Monte Carlo method,
it is desirable to determine the scattering type before we
determine the local velocity $\v u$ of the scatterer. 

We present a more quantitative argument about the preceding remarks.
Under the condition that a given photon is scattered off
by an atom located at a position $s$, the local velocity component
$u$ along the direction $\v k_i$ is simply given by
$$ 
f(u) \propto {e^{-u^2} \over  (u-b)^2 + a^2 }, \eqno(14)
$$
where $b=s-x$.  The normalization condition is used to get
$$ f(u) = {e^{-u^2} \over  (u-b)^2 + a^2 } 
\left[{\pi \over a} H(a,b)\right]^{-1}.
\eqno(15)
$$
Here, the Voigt function $H(a,b)$ is evaluated by a series expansion in $a$,
\ie,
$$
H(a,b) = H_0(a,b) + a H_1(a,b) +a^2 H_2(a,b) + a^3 H_3(a,b) + \cdots,
\eqno(16)
$$
where $H_n(a,b),~{n=0,1,2,3}$ are tabulated by Gray (1992).

Because of the smallness of $a$, the function $f$ has
a sharp peak around $u\approx b$, for which the scattering is resonant.
Therefore, the probability $P_r$ that a given scattering is resonant 
is approximately given by

$$ 
P_r  \simeq \int_{-\infty}^{\infty} d(\D u)
{e^{-b^2} \over  (\D u)^2 + a^2 }\left[{\pi \over a} H(a,b)\right]^{-1}
 = {e^{-b^2} \over H(a,b)}.
\eqno(17)
$$
The probability that a scattering occurs in the damping wing is

$$
P_{nr} = 1 - P_r.
\eqno(18)
$$

In the code we determine the scattering type 
in accordance with the scattering type probabilities $P_r$ and $P_{nr}$.
If a scattering is chosen to be resonant, then we set $u=b$.
Otherwise, the scattering occurs in the damping wing, and $u$ is chosen 
according to the velocity probability distribution given by Eq.~15.

We give the propagation direction $\v k_f$ of a scattered photon
in accordance with the isotropic phase function.
The scattered velocity component $v_\perp$
perpendicular to the initial direction $\v k_i$
on the plane spanned by $\v k_i$ and $\v k_f$ is also governed by
a Maxwell-Boltzman velocity distribution, which is numerically obtained
using the subroutine ${\it gasdev}()$ suggested by Press \etal (1989).
The contribution $\D x$ of the perpendicular velocity component 
$v_\perp$ to the 
frequency shift along the direction of $\v k_f$ is obviously
$$\D x = v_\perp [1-(\v k_i \cdot \v k_f )^2]^{1/2}/c. \eqno(19)$$

Therefore, the frequency shift $x_f$ of the scattered photon is given by
$$
x_f = x_i - u + u(\v k_i \cdot {\v k_f} ) 
    + v_\perp [1-(\v k_i \cdot \v k_f )^2]^{1/2},   \eqno(20)
$$
where $x_i$ is the frequency shift of the incident photon.

In each scattering event
the position of the scattered photon is checked and if it is
out of the medium we collect this photon according to
its frequency and escaping direction.
The whole procedure is repeated to collect typically 
about $10^3$ photons in each frequency bin.

\sec{3. Results}

\subsec{3.1 Escape Probability}

In Fig.~2 we show the emergent \Lya profile from a thick expanding medium. 
The horizontal axis represents $\Delta\lambda/\Delta\lambda_D=-x$,
and the vertical axis stands for the relative flux. Note that the
wavelength devitaion is the negative of the frequency deviation and
$\Delta\lambda_D$ is the Doppler wavelength.
The emission source lying in the center of the scattering medium is
assumed to be given by a Gaussian profile proportional to 
$e^{-(x/2\sigma_x)^2}$, 
where the width $\sigma_x$ is set to be 5. 
The use of a Gaussian profile 
may be justified by the observed Gaussian line profiles
for other emission lines such as $\rm H\alpha$ or [O III]
(Forbes \etal 1996, Legrand \etal 1997).
We choose $v_{th}=10~\km~\s^{-1}$, $s_{max}=10$, which corresponds to
$\D V_{bulk}=100~\km~\s^{-1}$. 
We also choose $\tau_0 = 10^4$ that leads to 
$N_{HI}=4.1\times 10^{18}~\cm^{-2}$ given the above choice of $s_{max}$.

The main features in the emergent profile in Fig.~2 include a primary 
peak in the red part and a much weaker secondary peak in the blue part.
One may understand this behavior qualitatively by considering the 
escape probability $P_{esc}$, with which a given photon escapes from 
the region to an observer. We note that the single-scattering escape 
probability is given by
$$P_{esc} = e^{-\tau_x},\eqno(21)$$ 
where the escaping optical depth $\tau_x$ is obtained  
from Eq.~10 setting $s=s_{max}$.

In Fig.~3 we show the single-scattering escaping optical depth 
and the escape probability.
The dashed line represents $\tau_x$, and the
$P_{esc}$ is shown by the dotted line.
The dot-dashed line is the input emission profile at the origin
given by a Gaussian and the solid line is the product of the initial
emission at the origin and the escape probability. The analysis has a
simple interpretation of the emergent profile without being scattered
from the source to the observer. 

Because the medium is not optically
thin, the escape probability analysis is a very poor approximation 
to the emergent profile which
is severely affected by photons scattered a large number of times. However,
the qualitative nature of the red asymmetric profile is obvious from the
figure.
The Sobolev optical depth $\tau_0 \gg 1$ for the incident photons with
frequency shift $x$ is satisfied in $-3 <x<s_{max}+3$. Therefore,
we expect a severe discrepancy in this regime.

In Fig.~4 we show the emergent profiles for the case where the incident
profile from the emission source is given by a flat
continuum. We use the same parameters as in Fig.~2.
The broad absorption feature is formed
in the frequency range of $-3 <x<s_{max}+3$ with
a sharp peak on the red side.

\subsec{3.2 Line Formation Mechanism}

When the static medium has 
a moderate line center optical depth $\tau_c<10^5$, 
the radiative transfer is dominated by the diffusion in frequency space,
and the spatial diffusion plays a rather minor role.
Hence each photon is scattered many times in the vicinity of its source
and gets a sufficient frequency shift just before it escapes, which
Adams (1972) described as ``a single longest flight.'' 
For a thicker medium with $\tau_c\gapprox10^5$, the damping wing scattering 
becomes important, and the photons emerge via the spatial diffusion.
This process was called ``a single longest excursion'' by Adams (1972).
He investigated these processes using a Monte Carlo method and 
showed that the mean scattering number $<N>$  before escape is given by
%%%
$$<N>\approx \cases{\tau_c\sqrt{\pi\ln\tau_c} & for $\tau_c <10^5$ \cr
1.5\pi^{1/2}\tau_c & for $\tau_c>10^5$. \cr}\eqno(22)$$

In the case of an expanding medium, using the Sobolev theory in a 
resonance shell having a width of velocity difference of order $v_{th}$, 
the effective optical depth is the Sobolev scattering optical
depth $\tau_0$  that is the total scattering optical depth in the
limit $\Delta V\rightarrow 0$. However, this hypothesis works only
when $\tau_0 \lapprox 10^3$, in which case a frequency diffusion
into the red wing leads to a direct escape from the scattering medium
without further scattering in the damping wing. 

When $\tau_0\ge 10^4$,
the damping wing scattering optical depth is not negligible and
the line photon can be captured with significant probability
despite the frequency diffusion into the red wing. The subsequent
scatterings tend to reduce the frequency deviation of the line
photon as the restoring force in the frequency space operates
(see Adams 1972). Therefore, the line photon is again resonantly
scattered after several wing scatterings and severe spatial
movement. In this case, the total scattering number is 
dominantly affected by the total column density of the scattering
medium, which is proportional to the product of $\tau_0$ and
$s_{max}$ according to Eq.~13.

Therefore, when $\tau_0=10^4$ and $s_{max}=10$,
the scattering number before escape is expected to be of order $10^5$ 
by regarding the resonance shell as a static medium with
a line center optical depth $\tau_c\sim\tau_0$ (Adams 1972).
It is checked that the scattering number increases almost linearly
with $\tau_0 s_{max}$.

%%%%%%%%%%%%%%%%%%%%%%%%%%%%%%%%%%%%%%%%%
In Fig.~5a is shown a typical scattering behavior of a photon
with $x \sim -5$.
The horizontal axis represents the scattering number.
In the bottom panel is shown the frequency variation of the photon.
In the middle panel, we show the radial distance of the photon from
the center of the sphere with radius $s=10$.
The scattering type is shown in the top panel in which
we denote a wing scattering by 0 and
a resonance scattering by 1.

Near the 360th scattering, the frequency of the photon deviates much 
from the center frequency, and several scatterings in
the damping wing follow. During this process, the photon propagates 
a large distance in real space and becomes resonant again by the effect 
of the restoring force (in frequency space) stated by Adams (1972).
When the photon escapes from the medium, it experiences several wing
scatterings, and the frequency of the photon falls on the red part.

In Fig.~5b we show a typical scattering behavior of a blue photon
with $x \sim 5$. Initially it experiences several damping wing 
scatterings and propagates easily with a large free path until
it hits the resonance zone.
Once it enters the resonance regime, the photon experiences the similar 
processes to those of the red photons described above.

On the other hand, majority of the extremely blue and the extremely red 
photons escape the medium suffering from 
at most several damping wing scatterings.

%%%%%%%%%%%%%%%%%%%%%%%%%%%%%%%%%%%%%%%%%
%In Fig.~6 we present resonantly scattered \Lya profiles for a number of
%initial Gaussian emission profiles having various line widths
%keeping the other parameters the same as in Fig.~2.
%Stronger secondary peaks are obtained as the widths
%of the input profile increase.
%This is explained by the fact that the secondary blue peak 
%is contributed by the frequency-redistributed photons initially
%emitted with frequencies $s_{max}-3 \lapprox x \lapprox s_{max}+3$
%and the photons with $x_i\gapprox s_{max}+3$ that emerge directly 
%without being scattered.
%Hence  the strength of the secondary peak sensitively depends on 
%the ratio of the width of the input \Lya profile to $s_{max}$.
%Observations of other emission lines having much smaller
%optical depths may provide complementary information on the initial 
%\Lya profile as suggested by Djorgovski \etal (1996).

\subsec{3.3 Dependence on Kinematics and Column Density}

In this subsection we investigate the dependence of the emergent
\Lya profile on the kinematics and the column density
of the scattering medium. Our numerical results are summarized
in Figs.~6 and 7a,b.

The kinematics of the scattering medium is characterized by
the parameter $s_{max}$ that measures the bulk velocity scale.
In Fig.~6 we show the emergent profiles for
$s_{max}=0.1, 1, 2, 5, 10$ with the Sobolev optical depth
$\tau_0=10^4$ and the width of the initial profile $\sigma_x=5$.
Here all the profiles have the same number of incident photons.
It is particularly notable that the slope at the blue edge of 
the primary peak does not change as $s_{max}$, which strongly
implies that this quantity can be very useful to put a 
strong constraint on $\tau_0$. This point is discussed
in more detail in the following section.

As $s_{max}$ gets smaller, the strength of the secondary peak
approaches that of the primary peak.
In the limit of $s_{max}\rightarrow 0$, we obtain a symmetric double peak
profile, and this is in agreement with the result obtained for the
case of a static medium by Adams(1972). A very turbulent medium 
or a static medium may provide a negligible $s_{max}$.

In Fig.~7a, we show the emergent profiles for moderate column densities,
$\tau_0=10^2,\ 10^3$ and $10^4$. Here, we fix $s_{max}=10$ and
use an initial Gaussian profile having a width of $\sigma_x=5$.
Because all the profiles have the same number of incident photons,
the profiles are normalized to have the same area.
The thick solid line indicates the emergent \Lya profile
for $\tau_0=10^4$ which corresponds to $N_{HI}=4.1\times 10^{18}~\cm^{-2}$,
the dotted line for $\tau_0=10^3$, and the dashed line for $\tau_0=10^2$.

The shapes of emergent profiles show a systematic change
with the Sobolev optical depth. The primary red peak recedes
to the red, as the column density increases. And the height of the primary
peak decreases and the width increases. As $\tau_0$ increases, the
range of frequency diffusion increases and therefore the width of the
primary red feature increases.

In the cases of high optical depths $\tau_0\ge 10^4$,
the profile of the emergent flux changes in a similar way.
In Fig.~7b we show the results for the cases $\tau_0=10^4,\ 10^5,\
$ and $10^6$. Up to $\tau_0 = 10^5$ the blue edges of the red primary
peaks appear to occur at the same location $x\simeq -2$. The slope
at the blue edge of the primary red peak decreases as the Sobolev
optical depth, that is, the overall profile changes 
from a sharp primary feature to a broad hump as $\tau_0$ increases. 
In contrast with the cases of Fig.~7a, the secondary blue peaks show
remarkable changes. As $\tau_0$ increases, the double peak profile
becomes more symmetric, which is obtained in the static limit. This
shows that the kinematics plays a minor role as the optical depth
increases.

For an extremely thick medium, $\tau_0\gapprox 10^6$,
the feature appears to have two broad humps outside the
absorption trough. The strength of the red part is now
almost similar to that of the blue part, which highly implies
that the anisotropy introduced by the kinematics has
little effect on the line formation. The enhanced width
of the feature can be a good measure of the column density of the
scattering medium.

It is noted that the strength ratio of the primary red part
to the secondary blue part is observationally important and
that several sets of ($\tau_0, s_{max}$) may give the same strength ratio.
In the following section we give a further discussion about observational
implications.
%Therefore, in order to lift this kind of degeneracy, we may use
%other observable quantities, such as the equivalent widths of the
%absorption

%These points are also important when one tries to fit
%the observed profile using a Voigt absorption profile,
%which is valid only in a static medium in thermal equilibrium.
%Lequeux \etal (1995) performed a detailed investigation
%on the \Lya profiles of the nearby blue compact dwarf galaxy,
%Haro 2. Using the Voigt profile fitting procedure,
%they estimated the H~I column density of an expanding shell around
%an H~II region in Haro 2 to be $N_{HI}=7.7\times10^{19}~\cm^{-2}$.

\subsec{3.4 Observational Implications}
We are primarily concerned with the quantities that can be observationally
measured and put meaningful and strong constraints on $\tau_0$ and
$s_{max}$ that specify the column density and the kinematics of the
expanding H~I medium.

We propose that the slope at the blue edge of the primary part
and the location of the primary peak can be one important
indicator of $\tau_0$ under the assumption that the spectroscopy
can be performed with enough resolution and signal to noise ratio.

In our work, the bin size for collecting the emergent photons
is $\Delta x=1$, that is, the bin size corresponds to the thermal
velocity $v_{th}=10 T_4^{1/2}~\km~\s^{-1}$. The normalization
is done by setting the total number of line photons to be 10000.
Therefore, the slope at the blue edge is defined by
the ratio of the photon number collected in the bin corresponding to
the maximum peak point to the number of bins between the location of
the blue edge and the maximum peak point of the primary red peak.

As shown in Fig.~7b, for $\tau_0=10^4,\ s_{max}=10$, the slope at the
blue edge of the primary part is given by
$$l={3800\ photons \over 1\ bin}, \eqno(23)$$
and similarly for $\tau_0=10^5$ it is given by
$$l={470\ photons \over 1\ bin}. \eqno(24)$$
It is particularly noted that these slopes are meaningful only
when the total number of \Lya photons throughout the feature is
normalized to 10000. It is also interesting that
$$l\le {20\ photons \over 1\ bin}. \eqno(25)$$
for $\tau_0=10^6$.

On the other hand, $s_{max}$ may be constrained by the quantities
including the peak-to-peak distance, or equivalently the width of the
absorption trough and the ratio of the strengths of the
primary peak and the secondary peak. However,
a good caution is needed because this quantity can be a poor indicator
when a huge column density is involved.

With lower quality spectroscopy we may use equivalent widths to
characterize the emission profile in a similar manner which Ringwald
\& Naylor (1997) used in their analysis of the lines from the
cataclysmic variable, BZ Camelopardalis.
If we measure observationally the equivalent widths $EW_1,\ EW_2,\ $
and $EW_{abs}$ corresponding to the primary red part, secondary blue
part and the absorption trough respectively, then the profile fitting
procedure will be reduced to minimize the differences of the
equivalent widths obtained from numerical simulations.

\sec{4. Application to DLA 2233+131}

DLAs are believed to be associated with a disk component of a normal 
spiral galaxy or a dwarf galaxy from the considerations 
of the similarity of the H~I column
density to a typical galactic value and the width of the
accompanying metal absorption systems amounting to a typical 
galactic rotational speed. There have been a few reports on the 
detection of the \Lya emissions in the spectra of DLAs
(\eg Djorgovski \etal 1996, 1997; M\"oller \& Warren 1993; 
Warren \& M\"oller 1996).

The first firm candidate is DLA 2233+131 reported by Djorgovski \etal (1996).
The emission feature is characterized by a sharp blue edge,
an absorption trough in the blue part, and a small secondary peak 
at around $5040~{\rm \AA}$. Djorgovski \etal (1996) ascribe 
the absorption feature to an ambient gas.

On the other hand, Lu \etal (1997) suggest that some \Lya forest clouds 
at lower redshifts may be responsible for the absorption feature in 
the blue part. However, as Djorgovski \etal (1996) and Lu \etal (1997)
point out, it is highly probable that DLA~2233+131 is a normal and 
isolated spiral galaxy. Therefore it will be a very rare coincidence 
if we have a \Lya emitting DLA system obscured by other intervening 
clouds at a smaller redshift.
It is more probable, therefore, that the absorption feature is formed
by the material surrounding the emission source.
Lu \etal (1997) also comment that the absorption
by either dust or the resonance scattering by neutral hydrogens
in the galaxy can significantly alter the energy distribution
of the escaping \Lya photons.

It is known that \Lya photons from a galaxy are mainly emitted
by the recombination processes in H~II regions.
Out of 8 nearby H~II galaxies, Kunth \etal (1996) find 4 galaxies
exhibiting a P-Cygni profile in the \Lya emission.
Subsequently, for one of such blue compact dwarf galaxies, Haro 2, 
Legrand \etal (1997) compared the observed P-Cygni \Lya profile 
with the profiles properly scaled 
from the observed $\rm H \alpha$ profile of Haro 2.
In their study it was found that 
the \Lya profile shows a significant excess in the red part.
They interpreted the red excess as the frequency redistributed flux
mainly caused by the line photons back-scattered from an expanding shell.
They conclude that the galaxies with a P-Cygni \Lya profile 
may possess superbubbles.

Furthermore, P-Cygni profiles are detected even at $3<z<3.5$ in the normal 
star-forming galaxies (Steidel \etal 1996b). 
The similar P-Cygni profiles are also detected in the spectra of galaxies
found in the Hubble Deep Field, where the redshifts are estimated to be
$z\sim 3$ (Lowenthal \etal 1997).

Therefore, we may draw a tentative conclusion that the P-Cygni type
absorption feature appearing in the \Lya emission of DLA~2233+131
strongly implies that the DLA system also contains similar superbubbles,
which are often found in nearby blue compact H~II galaxies.

From the width of the absorption trough we estimate 
the bulk velocity of the expanding shell in DLA~2233+131
to be $\D V_{bulk} \approx 
200~\km~\s^{-1}$ (see Fig.~2 in Djorgovski \etal 1996).
It is observationally challenging to measure the local thermal velocity
$v_{th}$ in the expanding H I region. The local velocity should also 
include the microturbulent motion, which may not be constrained 
observationally. Therefore, we assume that $v_{th}$ is in the range 
of $1-20~\km~\s^{-1}$ and adopt $v_{th}\approx 10~\km~\s^{-1}$ in this study.

The slope at the blue edge of the primary peak discussed in the previous
section is computed in the case of DLA~2233+131 observed by
Djorgovski \etal (1996). According to them, the observed \Lya line flux
$F_{1216}$ is
$$F_{1216}=(6.4\pm 1.2)\times 10^{-17}~\erg~\cm^{-2}~\s^{-1}.\eqno(26)$$
and the bin size is given by $\Delta\lambda\sim 0.8~{\rm\AA }$. The
maximum specific flux is $F_\nu\sim 8.5~\mu\Jy$, which is converted to
$F_\lambda \simeq 1.0\times 10^{-17}~\erg~\cm^{-2}~\s^{-1}~{\rm\AA}^{-1}$.
The number of bins between the
blue edge to the maximum peak point is 6. Using our normalization
scheme the slope is computed as
$$l_{2233+131}={10^4\over F_{1216}}{F_\lambda\Delta\lambda \over 6\ bins}
\sim 250 \eqno(27)$$
This result highly implies that the Sobolev optical depth $\tau_0$
for the outflowing H~I in DLA~2233+131 is a few times $10^5$, which
is converted to the column density $N_{HI}\sim 10^{20}~\cm^{-2}$
assuming that $s_{max}$ does not greatly exceed 50.
From these values and the overall 
shape of the profile, we exclude the possibility that 
$N_{HI} > 5 \times 10^{20}~\cm^{-2}$ or $\tau_0 > 10^6$. 

Assuming a standard Friedman cosmology 
with $\rm H_0=75~\km~\s^{-1}~Mpc^{-1}$
and $\Omega_0=0.2$, the angular diameter distance to the DLA
is $1.96\times 10^{28}~\cm$,
and the luminosity distance $r_L=8.14\times 10^{28}~\cm$.
We suppose that the H II region is a sphere of radius $R$.
Then the observed \Lya line flux is calculated by 
$$F_{1216} = {(h\nu_0)\alpha_B (n_en_p) \over r_L^2} ~V_{HII} ,
\eqno(28)$$
where $n_e$ is the electron density, $n_p$ is the proton density, 
$V_{HII}$ is the volume of the H II region, and the case B recombination 
coefficient $\alpha_B=2.59\times 10^{-13}~\cm^3~\s^{-1}$ 
for $T =10^4~\K$ (Osterbrock 1989).
From Eqs.~26, 28, we obtain the electron density $n_e$,
$$n_e \simeq 1~\left({1~\kpc \over R}\right)^3~\cm^{-3}. 
\eqno(29)$$
We extrapolate the correlation between velocity dispersion 
and radius of giant H~II regions (Terlevich \etal 1981) and
use the velocity dispersion $\sim 300~\km~\s^{-1}$
of the H~II region in DLA~2233+131 to 
estimate its size $R\sim 1~\kpc$. If we 
assume that the emission region is fully ionized, then
we get $n_e \simeq 1~\cm^{-3}$, which leads to the total mass
of the shell
$$M_{HII}\simeq 10^8 \msun. \eqno(30)$$

Because one O5 star typically generates UV photons at a rate of
$5\times 10^{49}$ photons $\s^{-1}$ (Spitzer 1978), 
it follows that the number of O5 stars needed to account for
the \Lya flux in DLA~2233+131 is $\sim 10^3$ under the assumption
that the H~II region is ionized purely by these O5 stars.
According to Kennicutt \etal (1989) the first-ranked H~II regions
found in nearby spirals and irregulars require the total mass 
of the ionizing stars
($>10\msun$) ranging $10^2-10^6~\msun$. From this, we may conclude
that the \Lya emission line from DLA~2233+131 is originated from
these first-ranked H~II regions.

\sec{4. Summary and Discussion}

We studied the line transfer in an optically thick 
and moving media including the damping wing scattering using 
a Monte Carlo method. 
The emergent profiles are characterized by the asymmetric double
peaks with a primary peak in the red part, a secondary peak
in the blue part, and an absorption trough in the blue.

The underlying mechanisms for the transfer and
the dependence of the emergent profiles on the H~I column density
and on the kinematics were investigated.
The primary red peak recedes to the red as $N_{HI}$ increases
and for thicker media with $\tau_0\ge10^6$, the \Lya line profile
becomes two separated broad humps. 
It is also found that as $s_{max}$ gets smaller,
symmetric double peak profiles are obtained, which is
regarded as the static limit where bulk motion is negligible.
We briefly discuss how to constrain the physical parameters $\tau_0$
and $s_{max}$ from an observed P-Cygni type profile.

We have applied this method to explain the P-Cygni type profile
in the \Lya emission of DLA~2233+131
assuming that it is caused by an expanding medium
with a high H~I column density. The slope at the blue edge of the
primary peak gives a strong constraint on $\tau_0$ and we find
that $\tau_0\simeq 10^5$.
Noting that the expanding bulk velocity $\D V_{bulk} \approx 200~\km~\s^{-1}$,
the column density responsible for the P-Cygni type feature is deduced to be
$N_{HI}\simeq 10^{20}~\cm^{-2}$.
We also present physical quantities regarding the H~II region and
deduced that $n_e \simeq 1~\cm^{-3}$, $R\simeq 1~\kpc$,
and $M_{HII}\simeq 10^8 \msun$, assuming near complete ionization.
About $10^3$ massive young stars are need to account for the
\Lya line flux of DLA~2233+131, is comparable with the
first-ranked H~II regions found in nearby spirals and irregulars.
\bigskip

%%%%%%%%%%%%%%%%%%%%%%%%%%
We briefly review the physical quantities of several
astronomical objects which may possess similar kinematic
properties to those of DLA~2233+131 discussed above.

The first example we consider is the H~I supershells and
the worms/chimneys in our Galaxy 
(Heiles 1979, 1984: Koo \etal 1991, 1992) and nearby galaxies.
An especially large H~I supershell is found
in the nearby normal spiral galaxy M101 (Kamphuis \etal 1991).
In M101 the measured line-of-sight expanding velocity component
of the supershell amounts to $\sim 50~\km~\s^{-1}$,
which is smaller than that of DLA~2233+131 by a factor of $\sim 5$.
The H~I column density of the supershell is estimated to be
$N_{HI} \simeq 2\times 10^{20}~\cm^{-2}$.

The second candidate is blue compact dwarf galaxies showing
similar P-Cygni type profiles in their \Lya emission
(Legrand \etal 1997, Lequeux \etal 1995, Kunth \etal 1996,
Puche \etal 1992). Kunth \etal (1996) found that 4 out of 8
observed H~II galaxies exhibit P-Cygni profiles.
One of them is Haro 2 whose H~I column density is
$2\times 10^{20}~\cm^{-2}$ and the expanding bulk velocity
$\sim 100~\km~\s^{-1}$ (Legrand \etal 1997).
As Legrand \etal suggested, the possibility for the H~II galaxy
to exhibit the \Lya emission or a broad absorption trough in their
spectra depends on geometrical properties of the expanding
shell, which include the angle between line-of-sight and the symmetry axis
of the shell.
%From the dearth of the observational data we may give an upper limit
%of the possibility of detecting \Lya from the H~II regoins to be
%$\sim 50$ percent and the the upper limit of the half-opening angle
%to be $ 60^\circ$ when we adopt the conical wind model
%of the emission line region proposed by Heckman \etal (1990).
It is noticeable that these blue compact galaxies are under
active star formation processes.

Currently, there are a few observational programs 
to discover primeval galaxies at $z>3$.

One is to search for star-forming galaxies like those found in 
$3<z<3.5$ by Lyman break characteristics 
(Steidel \etal 1996a, 1996b, 1997; Pettini \etal 1997).
The young galaxies detected in this way exhibit the similar P-Cygni type 
profiles to that found in DLA~2233+131. According to a deeper and 
higher resolution imaging (Giavalisco \etal 1996), these objects
appear to be compact spheroidal
cores often surrounded by low surface brightness nebulosities.
Assuming the standard cosmology of
$\Omega _0 = 1$ and $\rm H_0 = 50~\km~\s^{-1}~\Mpc^{-1}$,
a typical size of the cores is a few kpc.
The authors suggested that these are bulges of the primeval galaxies
which are precursors of nearby normal galaxies.

Another program is provided by gravitational lenses. This method is 
now becomes a powerful tool,
because the magnification of gravitational lens enables us
to detect more distant objects than other method does.
For example, the similar objects showing P-Cygni type profiles
are also found in the Hubble Deep Field at $z\simeq 3$
(Lowenthal \etal 1997). Trager \etal (1997) and Frye \etal (1997) also
found a very distant objects at $z\sim 4$.

It is remarkable that these objects show P-Cygni type
\Lya profiles. Especially the galaxy discovered by Frye \etal (1997)
shows several bright spots in the lensed image.
A particularly notable point is that only one of them (\ie N4 spot) 
shows a P-Cygni type \Lya line profile whereas others show damped 
\Lya absorption profiles (Bunker \etal 1997).
This may tell us that there are several well-localized H II regions in
the galaxy and only part of them shows a P-Cygni type \Lya emission, 
while the \Lya photons of others are
shielded by an optically thick medium along the line of sight.
Analogous situations are provided by nearby starburst dwarf
galaxies or supershells, and we suggest
that these starburst dwarf galaxies are the very counterparts
of the above three kinds of primeval galaxy candidates
including the DLA candidates.

Such outflowing media seen in extragalactic objects are often
accompanied by the sites associated with active star formation
(Reach \etal 1993).
Because DLA~2233+131 and other remote galaxies are believed to be
prototype of present galaxies,
it is very interesting that there may exist
primeval galaxies at $3<z<5$ showing starbursts, which also
gives constraints on the cosmological models for the structure formation.
Moreover this speculation supports the picture that star
formations are continuous processes from $z=0$ to $z\simeq 5$.

%%%%%%%%%%%%%%%%%%%%%%%%%%%%%
We propose complementary observations of metal lines in
the spectrum of DLA 2233+131 in order to provide
independent constraints on the relevant physical quantities.
We also propose the further imaging of DLA~2233+131
by the Hubble Space Telescope which may provide
independent and important clues to the identity of the system.

In a subsequent paper, we will investigate in more detail
the detectability of the \Lya emission in these star forming
objects, which is thought to be related with the geometrical factors
such as inclination associated with the shielding of the dust lane.

In numerical approach it will be desirable
to develop a more flexible code to treat inhomogeneous media
with an arbitrary velocity field and various shapes
to secure the constraints on the characteristic physical quantities
such as the column density and the bulk flow scale.

%%%%%%%========

\sec{Acknowledgements}
HWL gratefully acknowledges the support from the Research 
Institute for Basic Sciences, Seoul National University.
We are grateful to Prof. B. C. Koo and K. T. Kim for kind
and helpful discussions. We also gratefully acknowledge the kind
comments of Roger Blandford.

%%%%%%%%%%%%%%%%%%%%%%%%%%%%%%%%%%%%%%%%%%%%%%
%Reference macros
%Assumes five parameters:
%{Authors}{2 digit year}{Journal/Book+ed+pub as macro}{Vol}{Page}
%Third parameter ``0'' signifies in preparation
%Fourth parameter ``0'' signifies a book with a page number
%Fourth parameter ``.'' signifies a book without a page number
%Fifth parameter ``0'' signifies in press
\newdimen\refindent
\refindent=0.5truecm
\def\ref#1#2#3#4#5{
 \let\qzero=0
 \let\qpoint=.
 \def\qthree{#3}
 \def\qfour{#4}
 \def\qfive{#5}
 \filbreak\hangindent\refindent\hangafter=1\noindent
 \if\qpoint\qfour
  %Book reference without page number
  {#1 19#2. #3.}
 \else
  \if\qzero\qthree
   %Reference in preparation
   {#1 19#2. In preparation}
  \else
   \if\qzero\qfour
    \if\qzero\qfive
     %Book reference in press
     {#1 19#2. #3 in press.}
    \else
     %Book reference with page number
     {#1 19#2. #3, p.#5.}
    \fi
   \else
    \if\qzero\qfive
     %Journal reference in press
     {#1 19#2. {#3}, in press.}
    \else
     %Journal reference
     {#1 19#2. {#3}, {#4}, #5.}
    \fi
   \fi
  \fi
 \fi
}

%Journal macros

\def\mn{MNRAS}

%Publisher macros

%Book macros

%End reference macros
%%%%%%%%%%%%%%%%%%%%%%%%%%%%%%%%%%%%%%%%%%%%%%
\vfill\eject

\sec{5. References}
\ref{Adams T.}{72}{ApJ}{174}{439}
\ref{Blandford R. \& Lee H.-W}{97}{\mn \ submitted}{.}{}\par
\ni Bunker A. J., Moustakas L. A., Davis M., Frye B. L., Broadhurst T. J.,
\& Spinrad H.,\par
 1997, "The Young Universe: Galaxy Formation and Evolution at
 Intermediate and\par
 High Redshift" (Rome Observatory, 29 Sept - 3 Oct 1997) \par
 eds. S. D'Odorico, A. Fontana and E. Giallongo, A.S.P. Conf. Ser \par
%\ref{Churchill C. W}{96}{astro-ph/9604127}{.}{}
\ref{Djorgovski S.G., Pahre M.A., Bechtold J., \& Elston R.}
    {96}{Nature}{382}{234} \par
\ni Djorgovski S.G., 1997, 
     Structure and Evolution of the IGM from QSO Absorption Line\par
     Systems, IAP Colloquium, eds. P. Petitjean and S. Charlot, in press
\ref{Forbes D. A., Phillips A. C., Koo D. C., \& Illingworth G. D.}
    {96}{ApJ}{462}{89}
\ref{Franx M., Illingworth G. D., Kelson D., van Dokkum P., \& Tran, K.}
    {97}{ApJ}{486}{75}
\ref{Frye B. \& Broadhurst T.}{97}{submitted to ApJL}{.}{}
\ref{Giavalisco M., Steidel C. C., \& Macchetto F. D.}{96}{ApJ}{470}{189}
\ref{Gould A. \& Weinberg D. H.}{96}{ApJ}{468}{462}\par
\ni Gray D. F. 1992, in 
  {The Observation and Analysis of Stellar Photospheres 2nd ed.}\par
  Cambridge Press, New York\par
\ni Haehnelt M., Steinmetz M., \& Rauch M. 1997, ApJ submitted, 
astro-ph/9706201\par
\ref{Harrington J. P.}{73}{ApJ}{135}{195}
%%\ref{Heckman T. M., Armus L., \& Miley G. K.}{90}{ApJS}{74}{833}
\ref{Heiles C.}{79}{ApJ}{229}{533}
\ref{Heiles C.}{84}{ApJS}{55}{585}
\ref{Hunstead R.W., Pettini M., \& Fletcher A.B.}{90}{ApJ}{356}{23}
\ref{Irwin J. A.}{95}{PASP}{107}{715}
\ref{Kamphuis J., Sancisi R., \& van der Hulst T.}{91}{AAL}{244}{29}
\ref{Kennicutt R. C. JR., Edgar B. K., Hodge P. W.}{89}{ApJ}{337}{761}\par
\ni Koo B.-C, Heiles C., \& Reach W. T. 1991, in 
    {The Interstellar Disk-Halo Connection}\par
    {in Galaxies}, IAU Symp. 144, 
    ed. H. Bloemen, Kluwer, Dordrecht, p.~165\par
\ref{Koo B.-C, Heiles C., \& Reach W. T.}{92}{ApJ}{390}{108}\par
\ni Kunth D., Lequeux J., Mas-Hesse J.M.,Terlevich E., \& Terlevich R. 
    1996, in {Starburst}\par
    {Activity in Galaxies} 
    proceedings by the RevMexAstroAstrofis. (ConfSeries), \par
    astro-ph/9612043\par
%\ref{Lanzetta, K. M., Bowen, D. V.}{92}{ApJ}{391}{48}
%\ref{Lee H.-W}{97}{\mn \ submitted}{.}{}
%\ref{Lee H.-W}{94}{\mn}{268}{49}
\ref{Lee H.-W \& Blandford R.}{97}{\mn}{288}{19}\par
\ni Legrand F., Kunth D., Mas-Hesse J.M.,\& Lequeux J. 1997, {AA}, 
    in press, astro-ph/9706109.\par
\ref{Lequeux J., Kunth D., Mas-Hesse J.M., \& Sargent W.L.W}{95}
    {AA}{301}{18}
\ref{Lowenthal J. D. \etal}{97}{ApJ}{481}{673}
%\ref{Lu L., Wolfe A. M., Turnshek D. A., \& Lanzetta K. M.}{93}{ApJS}{84}{1}
\ref{Lu L., Sargent W. L. W. \& Barlow T. A.}{97}{ApJ}{484}{131}
%\ref{Lu L., Sargent W. L. W., Barlow T. A., Churchill C. W., Vogt S. S.}
%    {96}{ApJS}{107}{475}
%\ref{McKeith C. D., Greve A., Downes D., \& Prada F.}{95}{AA}{293}{703}
\ni  Mihalas, D. 1978, Stellar Atmospheres, W. H. Freeman and Company,
San Fransisco\par
\ref{M\"oller P. \& Warren S. J.}{93}{AA}{270}{43}
%%\ref{O'Connell R. W., Gallagher III J. S., Hunter D. A., \& Colley W. N.}
%%    {95}{ApJL}{446}{1}
\ref{Osterbrock D. E.}{62}{ApJ}{135}{195}\par
\ni Osterbrock D. E. 1989, Astrophysics of Gaseous Nebulae
    and Active Galactic Nuclei,\par
    University Science Books, California\par
%\ni Park C \& Kim J. 1997, submitted to ApJ, SNU-ASTRO-005.\par
\ni Pettini M., Hunstead R.W., King D.L., \& Smith L.J.,
    1995, astro-ph/9502076.\par
\ni Pettini M., Steidel C. C., Adelberger K. L., Kellogg M., Dickinson M., 
    \& Giavalisco M., 1997, astro-ph/9708117.\par
\ni Press W. H., Flannery B. P., Teukolsky S. A., \& Vetterling W. T.
    1989, {Numerical Recipes},\par
    Cambridge Press, New York\par
\ref{Prochaska J. X. \& Wolfe A. M.}{97}{AAS}{190}{4704}\par
\ref{Puche D., Westpfahl D., Brinks E.,\& Roy J,-R.}{92}{AJ}{103}{1841}\par
\ni Reach W. T., Heiles C., \& Koo B,-C. 1993, 
    in {\it AIP Conf. Proc. 278, Back to Galaxy},\par
    ed. S. S. Holt and F. Verter (New York, AIP), p.67.\par
\ni Ringwald F. A. \& Naylor T., 1997, AJ accepted. 
\ref{Rybicki, G. B. \& Hummer, D. G.}{78}{ApJ}{219}{654}
\ni Rybicki G. B \& Lightman A. P. 1979, { Radiative Processes in 
    Astrophysics}, John Wiley \& Sons, New York\par
%\ref{Scarrott S. M., Eaton N., \& Axon D. J.}{91}{MNRAS}{252}{12p}
\ref{Sengupta S.}{94}{MNRAS}{269}{265}\par
\ni Sobolev, V. 1960, 
Moving Envelopes of Stars. Harvard Univ. Press, Cambridge, Mass.\par
\ni Spitzer L., 1978, Physical Processes in the Interstellar Medium,
John Wiley \& Sons,\par
     New York\par
\ref{Steidel C. C., Giavalisco M., Pettini M., Dickinson M., 
     \& Adelberger K. L.} {96b}{ApJL}{462}{17}
\ref{Steidel C. C., Giavalisco M., Dickinson M., \& Adelberger K. L.}
    {96a}{AJ}{112}{352}
\ref{Steidel C. C., Adelberger K. L., Dickinson M., Giavalisco M., 
Pettini M., \& Kellogg M.}{97}{ApJ in press}{.}{}
\ref{Terlevich R. \& Melnick J.}{81}{MNRAS}{195}{839}
\ref{Trager S. C., Faber S. M., Dressler A., \& Oemler A.}{97}{ApJ}{485}{92}
\ref{Warren S. J. \& M\"oller P.}{96}{AA}{311}{25}
%\ref{Wolfe A. M., Turnshek D. A., Lanzetta K. M. \& Lu L.}
%{93}{ApJ}{404}{480}

\vfill
\eject

\sec{FIGURE CAPTION}

Fig.~1 \par
\ni A surface plot of the $s(x,\tau)$ table. $x\equiv\Delta\nu/\Delta\nu_D$
and $TAU=\tau$ represents the optical depth. 
The plateau on the right top side corresponds to $s=2s_{max}+1$.
\bigskip

Fig.~2 \par
\ni The emergent \Lya profile from a thick expanding medium.
The horizontal axis represents $\Delta\lambda/\Delta\lambda_D$
and the vertical axis for the relative flux.
The emission source is assumed to be given by a Gaussian profile 
$\propto e^{-(x/2\sigma_x)^2}$, where the width $\sigma_x$ is set to be 5.
We choose $s_{max}=10$ and $\tau_0 = 10^4$.
The emergent profile is represented by the solid line
and the dotted line shows the initial Gaussian profile.
\bigskip

Fig.~3 \par
\ni The escaping optical depth and the single-scattering escape probability.
The dashed line represents the optical depth $\tau_x$, 
and the escape probability $P_{esc}\equiv e^{-\tau_x}$ is shown by 
the dotted line.
The dot-dashed line is the input emission profile at the origin
given by a Gaussian with the width $\sigma_x=10$.
The solid line is the product 
of the initial emission at the origin and the escape probability.
\bigskip

Fig.~4 \par
\ni The emergent profiles for a flat incident continuum.
The parameters are the same as in Fig.~2.
\bigskip

Fig.~5a \par
\ni A typical scattering behavior of a photon with $x \sim -5$
The horizontal axis stands for the scattering number. In the bottom
frequency deviation is shown, and in the top panel the scattering type
is given, where 0 stands for a wing scattering and 1 for a resonant
scattering. In the middle panel the radial distance of the photon
from the center of the scattering sphere is shown.
\bigskip

Fig.~5b \par
\ni Another typical scattering behavior of a photon with
$x \sim 5$. See the text for more detail. \par
\bigskip

Fig.~6 \par
\ni The emergent \Lya profiles for various bulk velocity scales
$s_{max}$ of the
scattering medium. We fix $\tau_0=10^4$ and the initial Gaussian
profile having a width of $\sigma_x= 5$ is used.
The thick solid line is for the case of $s_{max}=0.5$, the dotted line
for $s_{max}=2$, the long dashed line for $s_{max}=1$, the dashed line
for $s_{max}=5$, and the dot-dashed line for $s_{max}=10$.
\bigskip

Fig.~7a \par
\ni The emergent \Lya profiles for various column densities of the 
scattering medium.
$s_{max}=10$ and the initial Gaussian profile having a width of
$\sigma_x=5$ is used. The thick solid line indicates 
the emergent \Lya profile for $\tau_0=10^4$ which corresponds
to $N_{HI}=4.1\times 10^{18}~\cm^{-2}$, the dotted line for $\tau_0=10^3$,
and the dashed line for $\tau_0=10^2$.
The small dips in the secondary
peaks are numerical artifacts caused by the truncated matter distribution.
\bigskip

Fig.~7b \par
\ni The emergent \Lya profiles for various column densities of the
scattering medium.
The thick solid line indicates
the emergent \Lya profile for $\tau_0=10^4$ which corresponds
to $N_{\rm HI}=4.1\times 10^{18}~\cm^{-2}$,
the dotted line for $\tau_0=10^5$,
and the dashed line for $\tau_0=10^6$.

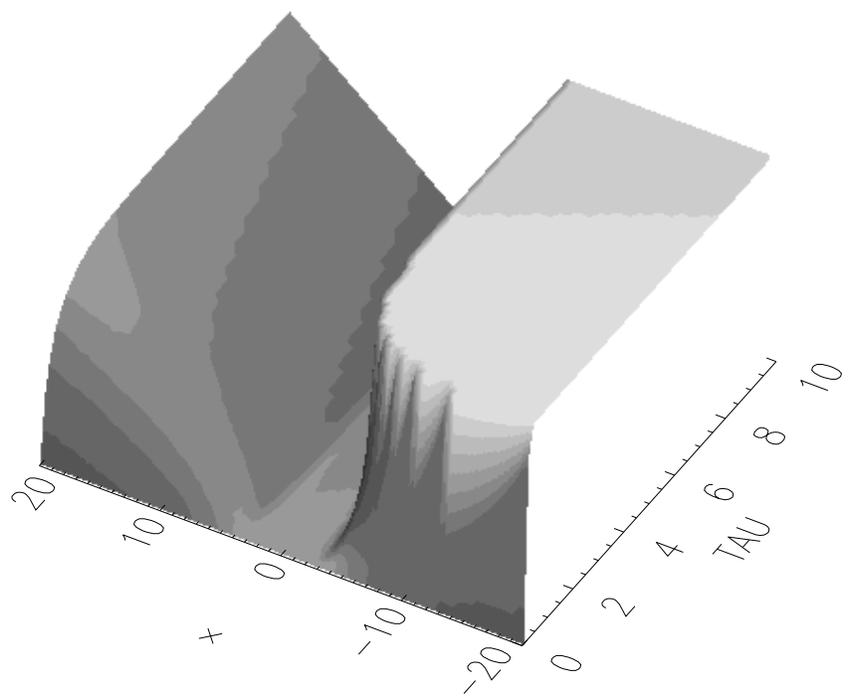
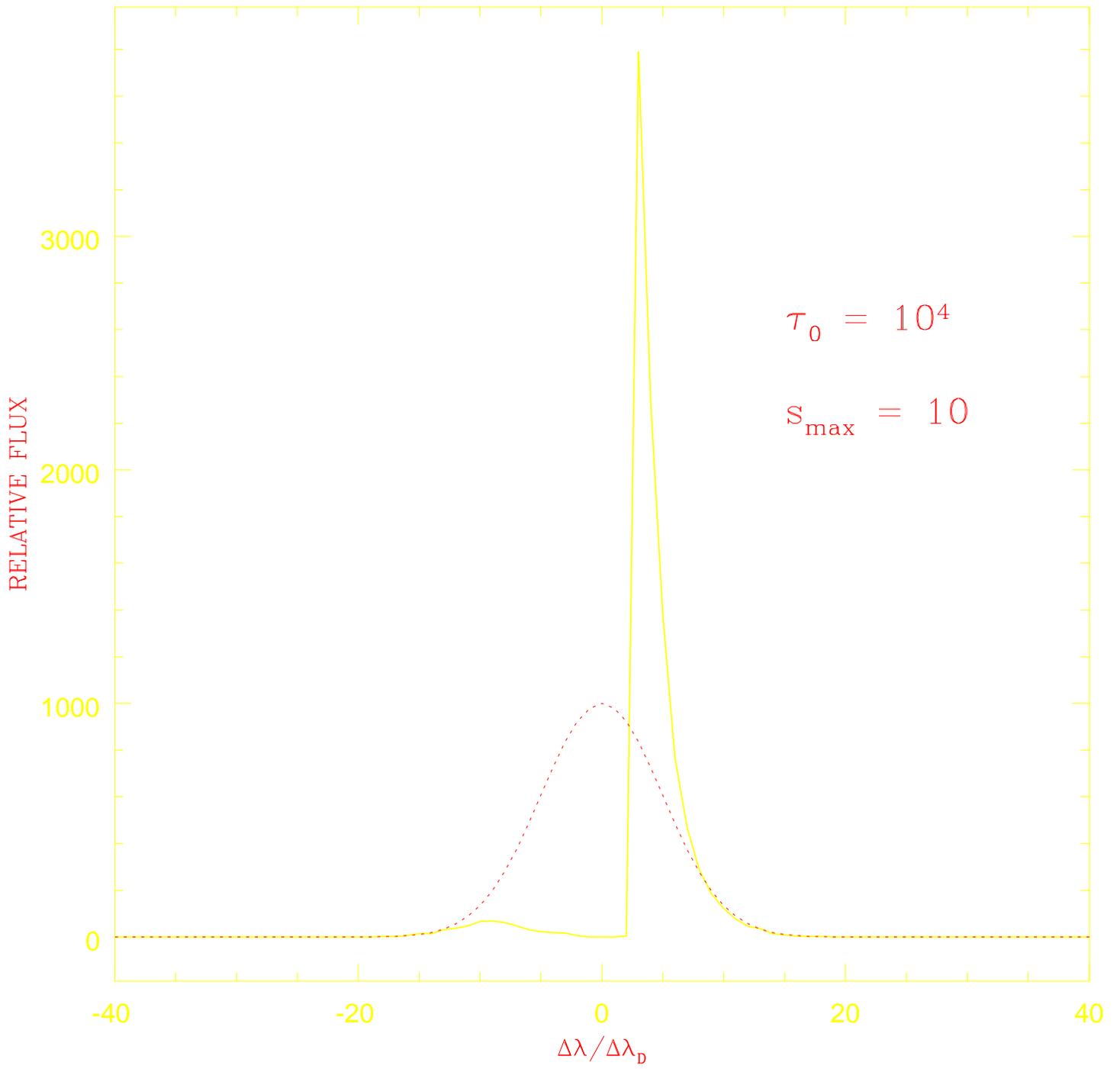
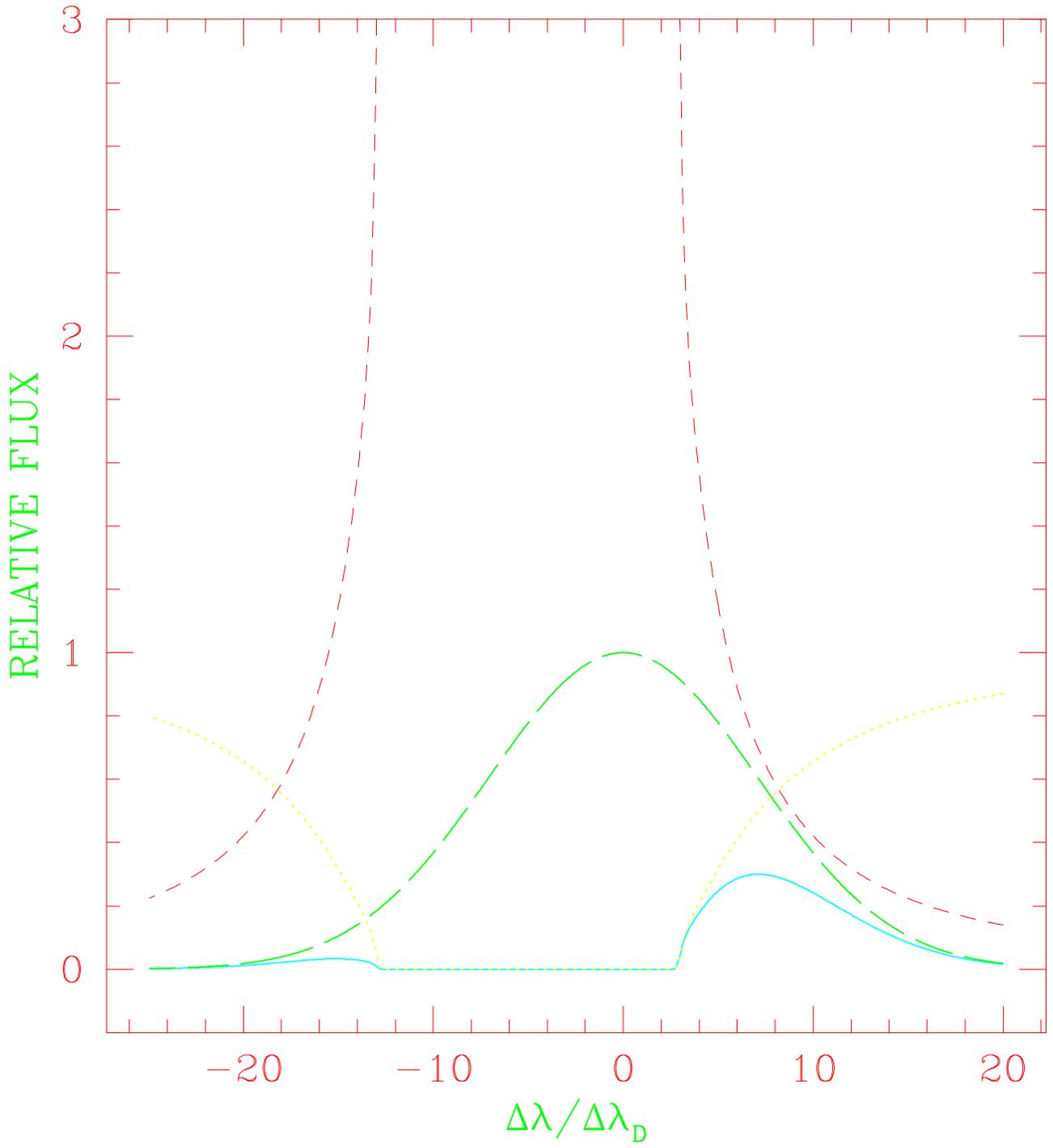
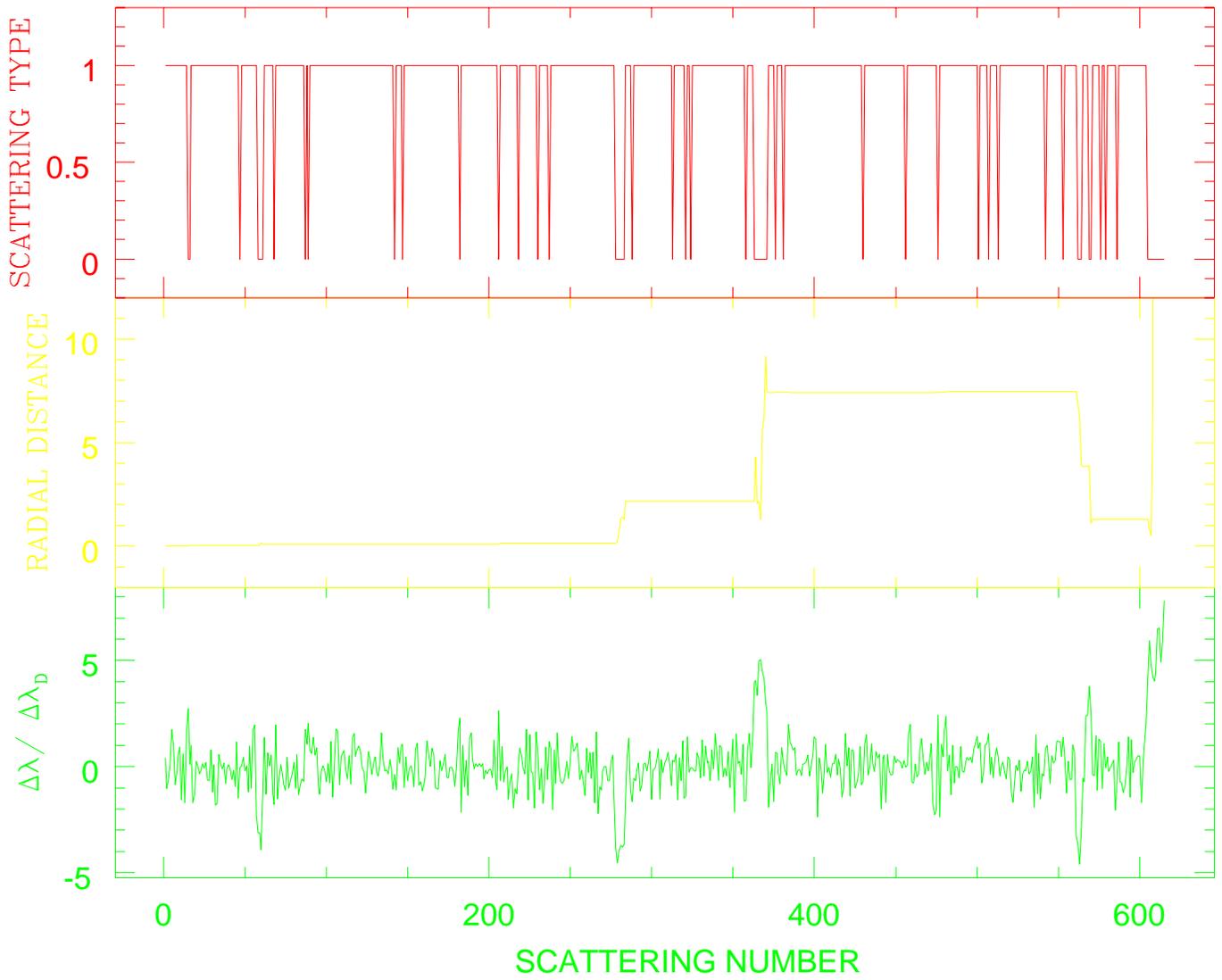
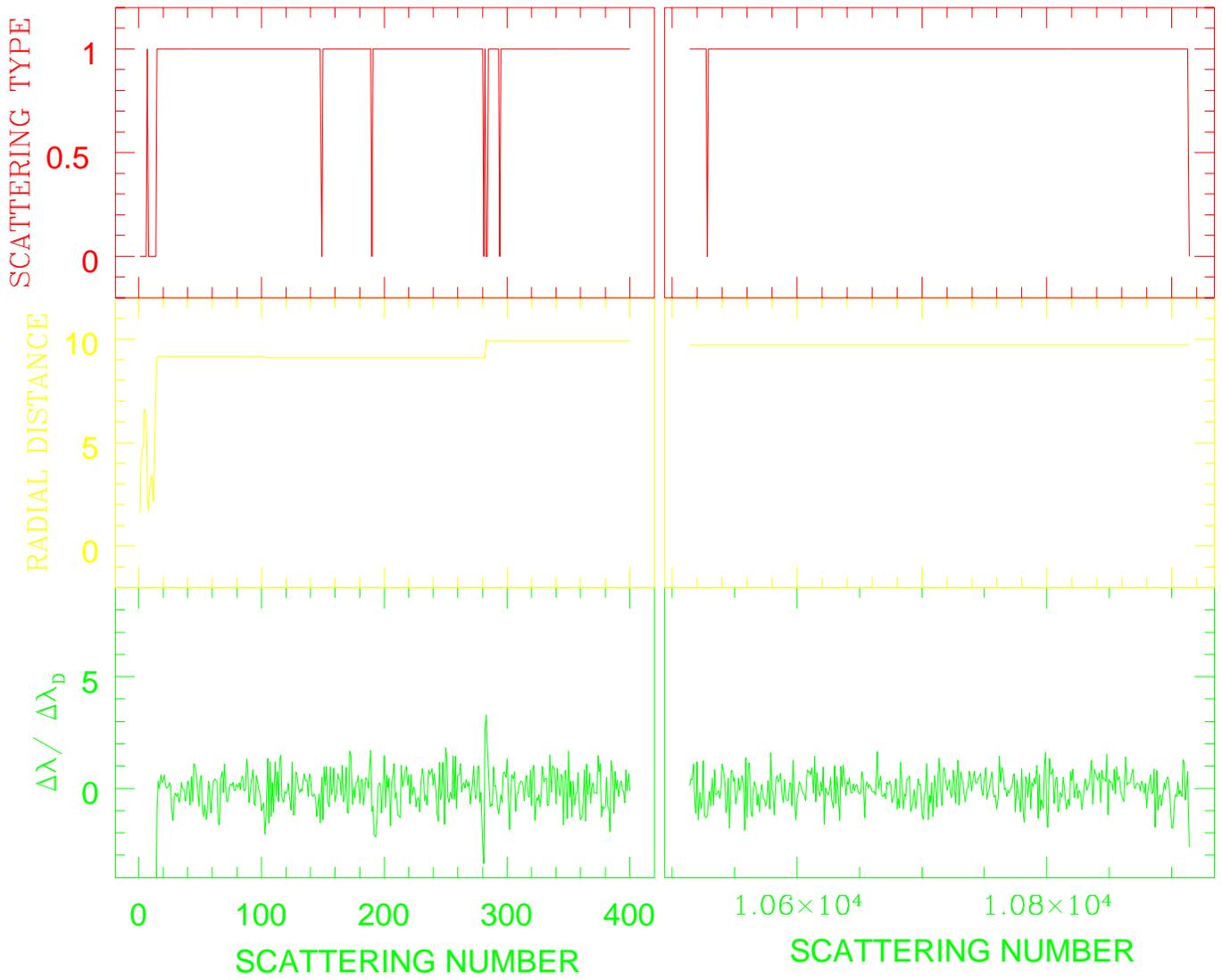
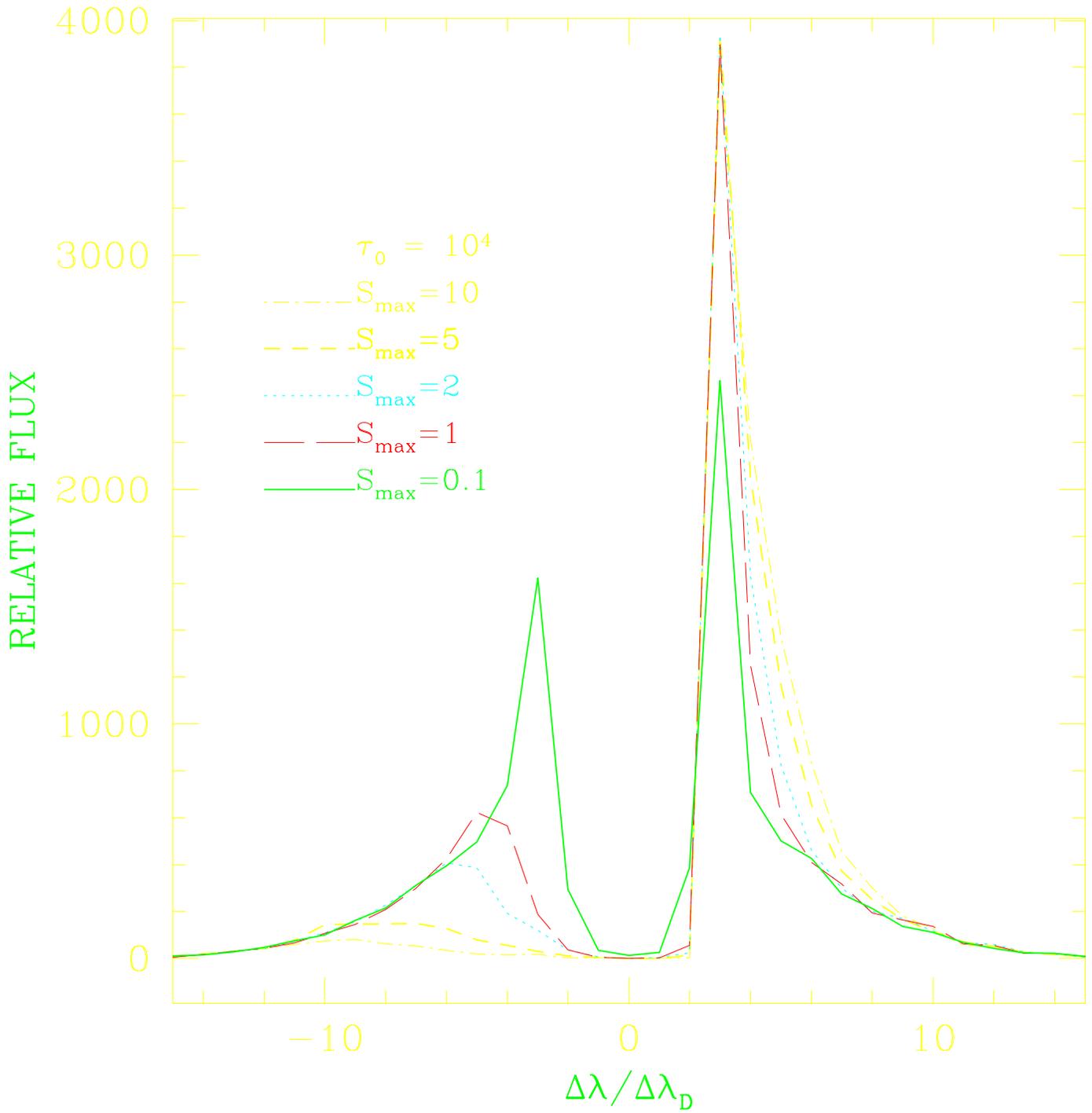
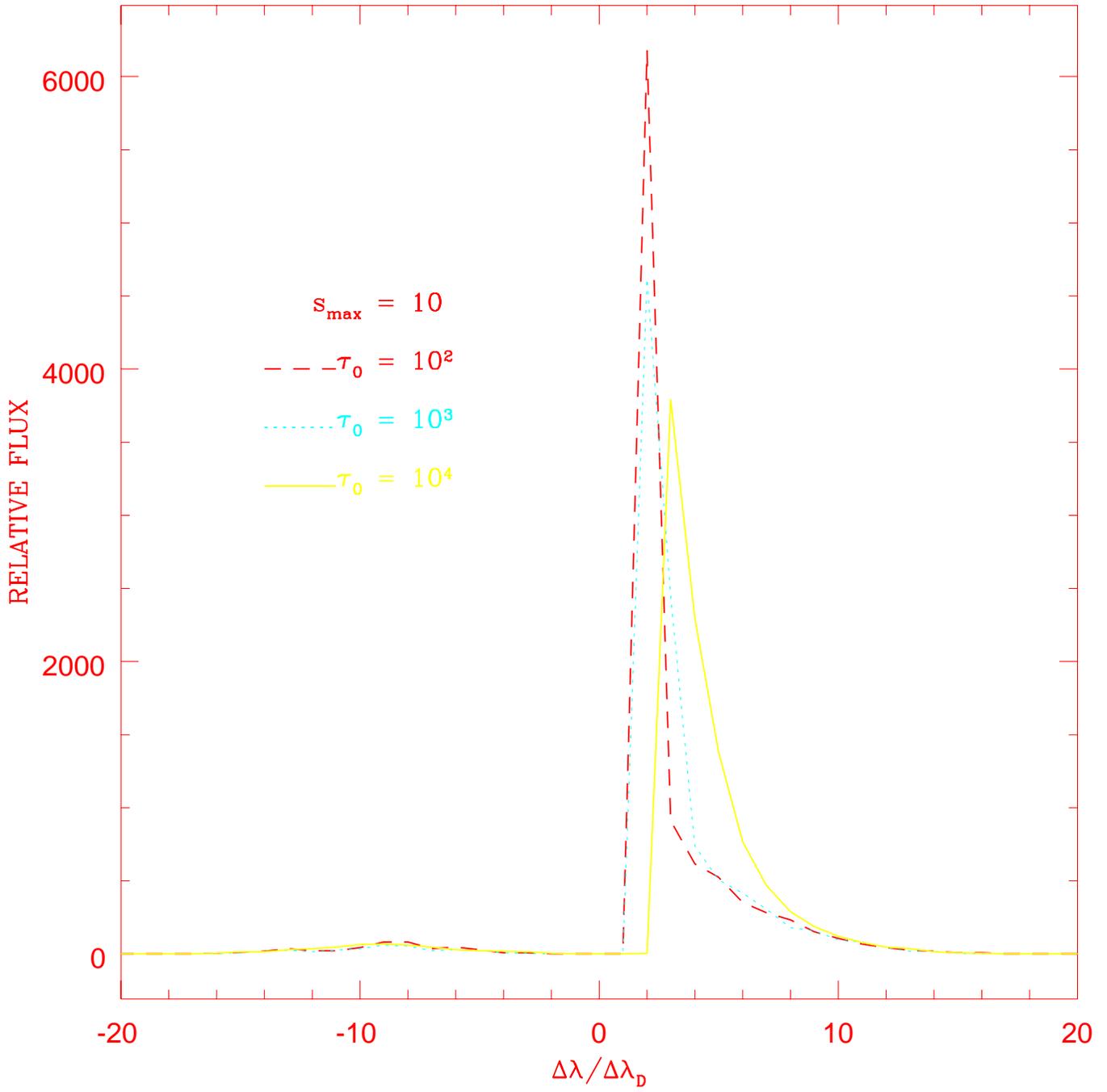
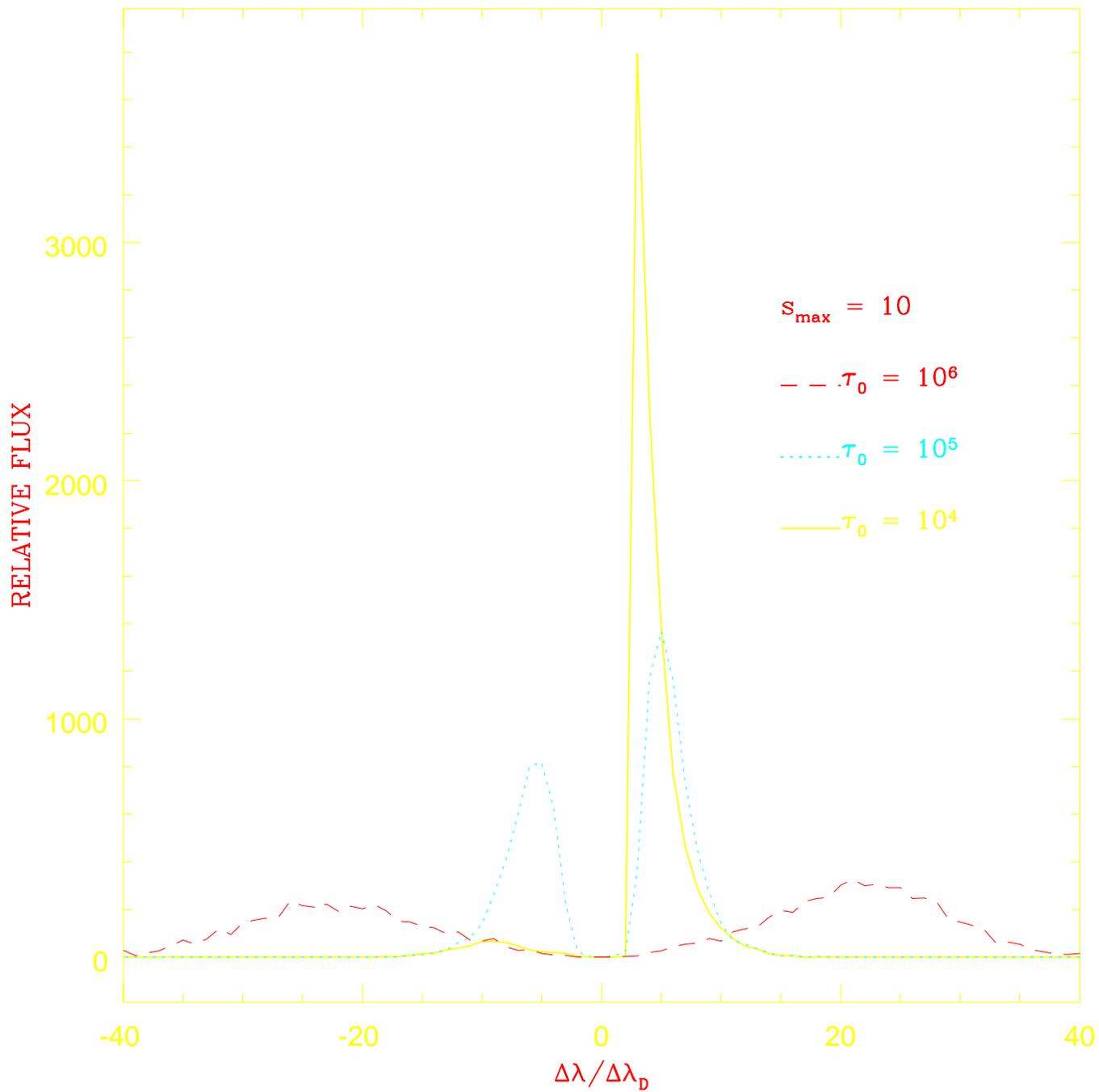
\end